\documentclass[preprint,flushrt]{aastex}
\usepackage{amsmath}
\usepackage{graphicx}
\usepackage{subfigure}
\usepackage{ulem}
\usepackage[hyphens]{url}

\shorttitle{TWA}
\shortauthors{Donaldson et al.}

\begin{document}

\title{New Parallaxes and a Convergence Analysis for the TW Hya Association}

\author{J.\ K.\ Donaldson\altaffilmark{1}, A.\ J.\ Weinberger\altaffilmark{1}, 
J.\ Gagn{\'e}\altaffilmark{1, 3}, J.\ K.\ Faherty\altaffilmark{1, 2, 4}, A.\ P.\ Boss\altaffilmark{1}, S.\ A.\ Keiser\altaffilmark{1}}
\affil{Department of Terrestrial Magnetism, Carnegie Institution of Washington, 5241 Broad Branch Rd NW, Washington, DC 20015, USA; \texttt{jdonaldson@carnegiescience.edu}}
\altaffiltext{2}{Department of Astrophysics, American Museum of Natural History, Central Park West at 79th Street, New York, NY 10034, USA}
\altaffiltext{3}{NASA Sagan Fellow}
\altaffiltext{4}{NASA Hubble Fellow}

\begin{abstract}
The TW Hya Association (TWA) is a nearby stellar association with an age of $\sim$\,5--10\,Myr. This is an important age for studying the late stages of star and planet formation.  We measure the parallaxes of 14 candidate members of TWA. That brings to 38 the total number of individual stars with fully measured kinematics, i.e.\ proper motion, radial velocity, and parallax, to describe their motions through the Galaxy.  We analyze these kinematics to search for convergence to a smaller volume in the past, but we find the association is never much more compact than it is at present. We show that it is difficult to measure traceback ages for associations such as TWA that have expected velocity dispersions of \mbox{1--2\,km\,s$^{-1}$} with typical measurement uncertainties. We also use our stellar distances and pre-main-sequence evolutionary tracks to find the average age of the association of  $7.9\pm1.0\,$Myr. Additionally, our parallax measurement of TWA 32 indicates it should be considered a bona fide member of TWA. Two new candidate members have high membership probabilities, and we assign them TWA numbers: TWA 45 for 2MASS J11592786--4510192 and TWA 46 for 2MASS J12354615--4115531.
\end{abstract}

\section{Introduction}

Nearby, young stellar associations ($\sim5$--10\,Myr) are the perfect place to study the last stages of the formation of stars and planets. Many young stars are surrounded by disks that are the sites of planet formation. Nearby disks, such as the one around TW Hydrae, provide ample opportunity to study this process in detail. But to put those observations in context, the ages of these stars must be accurately determined.

Ages of young stellar associations can be more accurately determined than those of single stars \citep{Zuckerman04,Torres08}. Assuming the stars in the association are all born at around the same time, the age of the association can be determined by combining information on the ages of individual stars.  Additionally, traceback analyses can be done to trace the kinematic history of the stars back to the place of their births.  

The TW Hya Association (TWA) is a nearby \citep[$\sim 50$\,pc;][]{Zuckerman04} association of stars that have been identified by their common proper motion \citep{Webb99,Sterzik99,Zuckerman01,Gizis02,Song03,Scholz05,Looper07, Shkolnik11}.  The age of TWA \citep[$10\pm3$\,Myr;][]{Bell15} has been the subject of much debate \citep{Mamajek05,Makarov05,delaReza06,Weinberger13,Ducourant14}.  A few members have young looking, gas-rich disks, while other members are diskless or have older looking debris disks \citep{Weinberger04,Low05,Riaz08,Plavchan09,Schneider12,Riviere13}.  

Distance is an important parameter in accurately determining age and membership of stars in TWA. Only four members of TWA had distances measured by Hipparcos \citep{Perryman97}.  More recently, parallaxes and proper motions of many more stars have been measured \citep{Gizis07,Teixeira08,Teixeira09,Weinberger13, Ducourant14}, allowing for more rigorous analyses of kinematics. \cite{Weinberger13} and \cite{Ducourant14}, however, disagreed over the traceback analysis; \cite{Weinberger13} claimed no convergence of the association, while \cite{Ducourant14} obtained a traceback age of $7.5\pm0.7$\,Myr. 

TWA candidates are usually identified through a combination of proper motion, spectral signatures of youth, and radial velocities. The hardest parameter to measure is the parallax. Stars can only truly be classified as members after their full kinematics have been measured. Only after membership is confirmed, do they form a coherent sample in which to study disk evolution or with which to calibrate youth indicators such as Li  and Na equivalent widths. In this paper, we present parallaxes and proper motions of 14 proposed TWA members. This sample expands upon the survey done by \cite{Weinberger13}, using the same instrument to double the number of measurements. We also include 7 parallaxes from the GAIA Data Release 1 catalog \citep{GAIAdr1}.

\section{Observations\label{sub:obs}}
\subsection{Astrometric Data}

We have observed 14 potential members of TWA with the CAPSCam instrument \citep{Boss09} at the 2.5 m du Pont Telescope at Las Campanas Observatory. A description of the instrument and data taking are given in \cite{Boss09} and \cite{Weinberger13}, and only a few salient points are given here. The instrument makes use of an independently readable subarray (the guide window; GW). For our observations the GW is placed on the relatively bright target star in the center of the full array and read out more rapidly than the full field (FF). The FF is integrated deeply to obtain a higher signal-to-noise ratio (S/N) on the reference stars. Table~\ref{tab:obs} lists the typical GW and FF integrations for each target. Observations for each target at each epoch were typically repeated 16 times, 4 times for 4 dither positions.

We operated in the no guide window shutter mode (NGWS), where the shutter remains open during the GW readout. This was done rather than using the GW shutter mode (GWS) to improve the efficiency of the observations.

Our targets were chosen to round out the survey of TWA members done by \cite{Weinberger13}.  The new targets include TWA 6, 8, and 10 \citep{Webb99}, TWA 17 and 18 \citep{Zuckerman01}, TWA 30 A \citep{Looper10}, TWA 31 and 32 \citep{Shkolnik11}, TWA 33 \citep{Schneider12}, 2MASS J10252092--4241539 \citep[2M1025--42;][]{Rodriguez11}, 2MASS J11592786--4510192 \citep[2M1159--45;][]{Rodriguez11}, 2MASS J13112902--4252418 \citep[2M1311--42;][]{Rodriguez11}, and 2MASS J12354615--4115531 \citep[2M1235--41;][]{Riaz06}. 

\section{Data Analysis\label{sub:data}}
\subsection{Astrometric Solution}

We combined all epochs of the observations to calculate an astrometric solution of position, proper motion, and parallax for all the targets. All targets were observed in at least 5 epochs, the dates of which are listed in Table~\ref{tab:obs}.  The dates of the observations are spread out in parallax factor so the astrometric solution can be well constrained.

The astrometric solution is calculated using ATPa, as described in \cite{Boss09} and \cite{Anglada12}. The calculation is an iterative process. One observation is chosen to be the template, and the positions of stars in the field are extracted. For each image, a set of reference stars are matched to the template, and their positions are averaged together for each epoch. Their apparent motion is then fed into an astrometric model to obtain positions, proper motions, and parallaxes for all stars. The reference frame is then adjusted to select only those stars with the smallest residuals, and the process is iterated a few more times.

\subsection{Zero-point correction}

Since the position of the target star is measured against the position of background stars, their motion must be taken into account as well. While more distant stars will have smaller parallactic motion, it can be non-zero and can introduce a bias into the parallax calculations.  Therefore, we must correct for the average parallax of the reference stars.

Photometric distances are found for the reference stars by fitting USNO $B1$-$B2$, $R2$ and $I$ and 2MASS $J$, $H$, and $K_\text{S}$ photometry with Kurucz stellar atmosphere models.  Dwarf stars with $T_\text{eff} < 3800$\,K are not used as references because the models are not considered reliable.  Giant stars are recognized by the small distances they produce when fit as dwarfs, and are refit as giants.  We then look at the average difference between the photometric parallax and the astrometric parallax as an estimate of the bias.  We subtract the bias from the relative parallaxes to get the absolute parallaxes of the target stars.  The relative parallaxes, zero-point corrections, and absolute parallaxes are listed in Table~\ref{tab:para}. More details on the zero-point correction can be found in \cite{Anglada12} and \cite{Weinberger13}.

\section{Traceback age\label{sub:traceback}}
One method to get the age of TWA is to trace back the locations of the stars to find the time when they were closest together. This should correspond to the time when the stars were born.

To trace back the stars in time, we need full kinematics of the stars. With our CAPSCAM data, we measure the parallax and proper motion of the stars. This combined with Radial Velocity (RV) data from the literature allows us to calculate the full $UVW$ motions of the stars. We combined our data with parallax, RV, and proper motion values from the literature, including new GAIA parallaxes of 7 TWA stars \citep{GAIAdr1}.  We calculated weighted averages of the independent measurements.  The values we adopted and their sources are listed in Table~\ref{tab:velocities}.

Once we have the velocities, we calculate the present position of the stars relative to the Sun and subtract the velocities multiplied by the time elapsed to get the positions of the stars back in time. We then calculate a radius of the cluster at each time by deriving the average distance of the stars from the center, defined by the mean location of all the stars. The traceback age is defined as the time at which the radius is at a minimum. 

\begin{figure}[t!]
\begin{center}
	\vspace{-4cm}\includegraphics[width=0.8\textwidth]{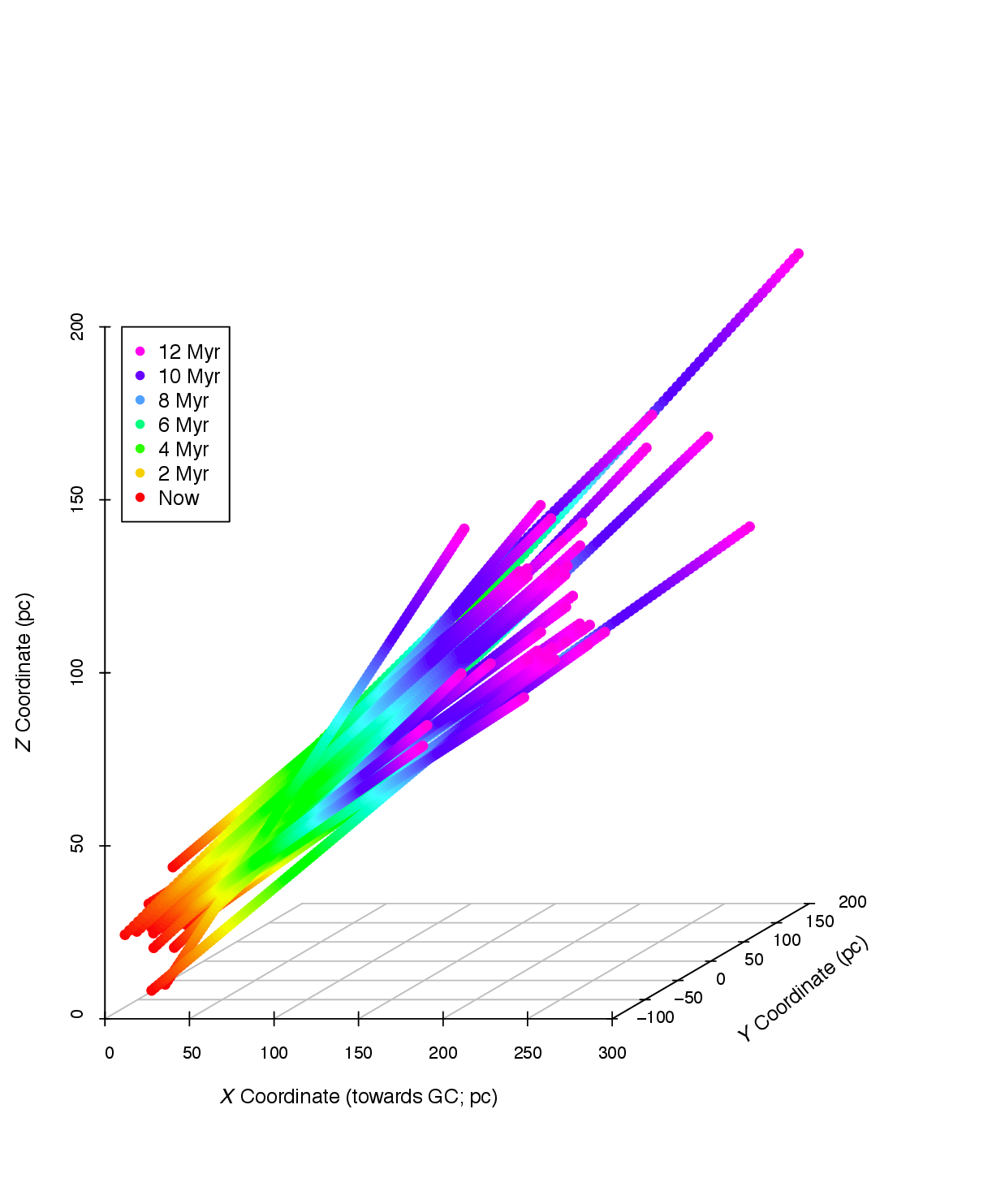}
	\caption{The galactic $XYZ$ positions of the TWA stars as a function of time.  The stars diverge after only 1.6\,Myr. See Section~4 for more details.}
	\label{fig:traceback}
\end{center}
\end{figure}

\begin{figure}[t!]
\begin{center}
	\includegraphics[width=\textwidth]{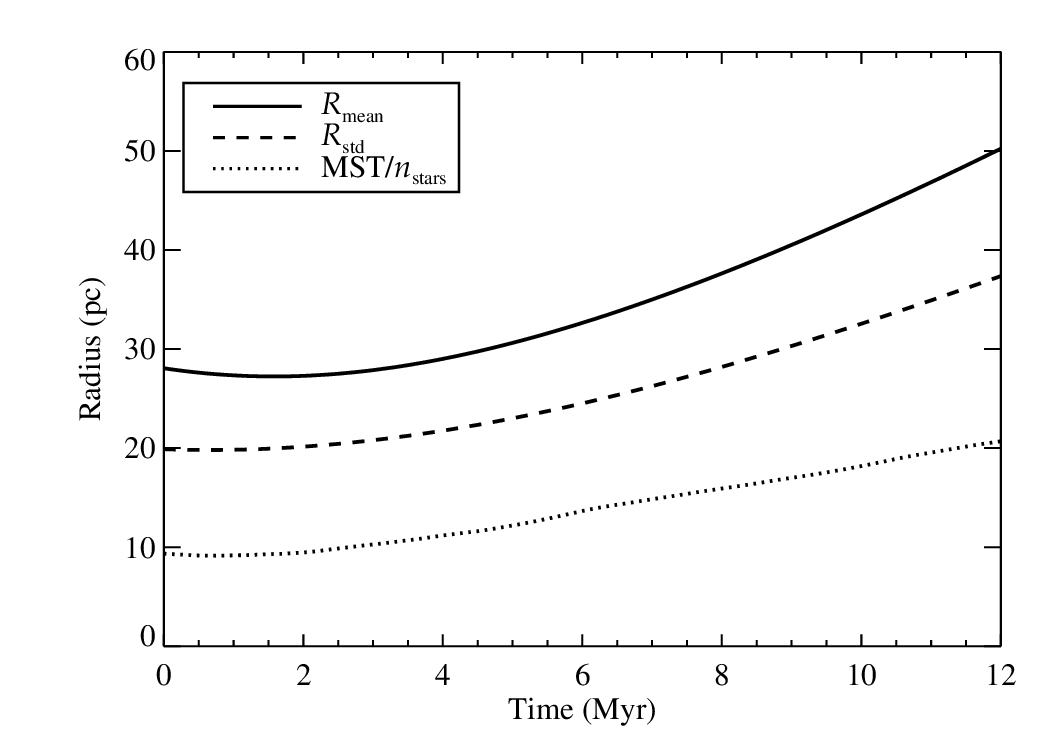}
	\caption{The radius of TWA as a function of time from the Monte Carlo traceback analysis.  The stars do not get much closer than they are today, and diverge after a traceback age of only 1.6\,Myr. The solid line represents $R_\text{mean}$, the mean distance of all the stars from the central point. The dashed line shows $R_\text{std}$, the standard deviation of the stars about the central point.  The dotted line is the Minimum Spanning Tree (MST) divided by the number of stars (38). See Section~4 for more details.}
	\label{fig:sigma}
\end{center}
\end{figure}

The uncertainties in our measurements will propagate forward to an uncertainty in our traceback age.  To account for this, we used a Monte Carlo (MC) method.  We ran 10,000 trials, where for each trial we included a random uncertainty to the parallax, proper motion, and RV values.  The random uncertainty is drawn from a normal distribution centered around zero with a standard deviation equal to the measured uncertainty for that star.  The final radius at each age is defined as the average over all 10,000 trials.  The uncertainty in the traceback age is taken as the standard deviation of the distribution of ages we get from each trial.
When we do this for all the stars, we get an age of $1.6\pm0.5\,$Myr, which is much younger than expected. Figure~\ref{fig:traceback} shows the positions of the stars as a function of time, and Figure~\ref{fig:sigma} shows the radius of TWA as a function of time. 

Since TWA is an elongated association, rather than spherical, we tried two other methods to calculate a characteristic scale size for the group. The first is the method of \cite{Ducourant14}, to define the radius by the standard deviation about the mean, $\sigma$, as $R_\text{std} = \frac{1}{3}\left(\sigma_x + \sigma_y + \sigma_z\right)$. The second is to use a Minimum Spanning Tree \citep[MST; e.g.\ ][]{Allison09}, the shortest path made by connecting all the stars with straight lines. Figure~\ref{fig:sigma} shows the radius of the association as a function of time using all three methods.  Both of these methods give a minimum size at a young age, 0.7\,Myr for $R_\text{std}$ and 0.8\,Myr for MST.  

\cite{Ducourant14} performed a traceback analysis and identified a core group of stars with velocities that were not outliers. The stars include TWA 1, 2, 3, 4, 7, 8A, 8B, 11A, 12, 13A, 13B, 14, 15A, 15B, 23, and 32.  
We do a traceback using only these stars, except TWA 3.  We exclude TWA 3 because it does not have a parallactic distance.  \cite{Ducourant14} used its convergent point distance, which depends on already having an understanding of what stars are members of TWA.  Using only the remaining 15 core stars, 
we get an age of $3.8\pm1.1\,$Myr for $R_\text{mean}$, 2.8\,Myr for $R_\text{std}$, and 3.9\,Myr for MST. 
The differences between our work and that of \cite{Ducourant14} is discussed in more detail in Section~\ref{sub:MC}.  In Section~6 we define a new core group based on our results. 

\begin{figure}[t!]
\begin{center}
	\includegraphics[width=\textwidth]{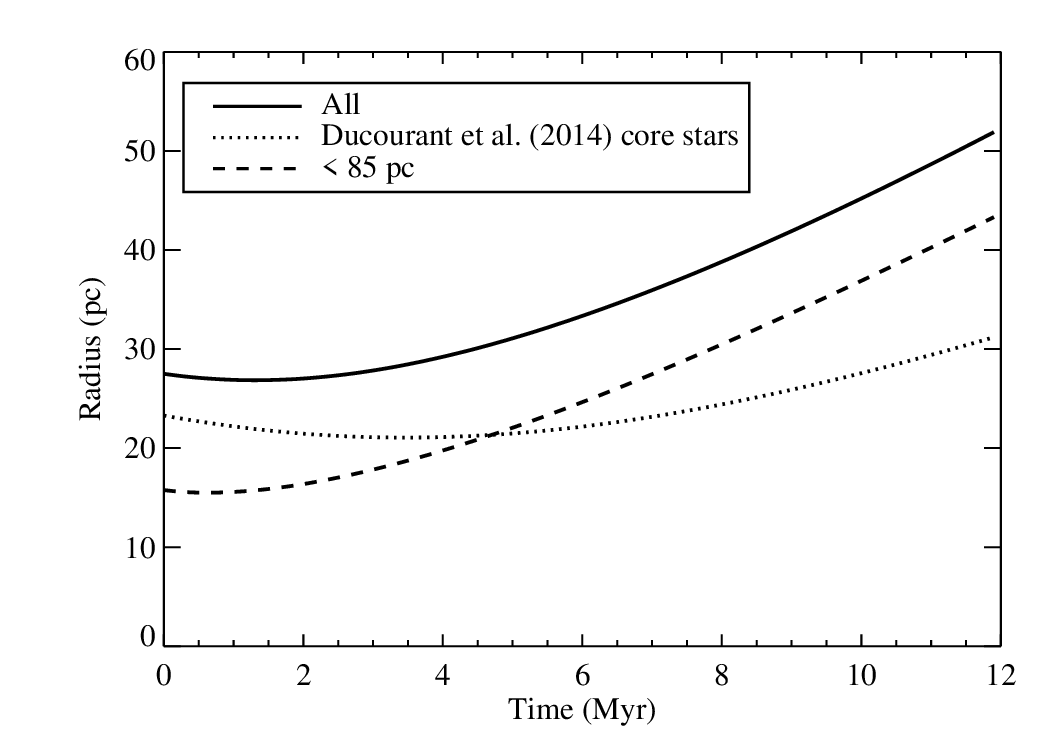}
	\caption{Radius of TWA as a function of time for three different subsamples of the Monte Carlo traceback analysis.  The solid line shows the full sample, 38 proposed TWA members with full kinematics.  The dashed line shows the subsample of 32 stars with distances less than 85\,pc. The dotted line represents the subsample of 15 core stars identified by \cite{Ducourant14}. All the subsamples diverge rapidly. The \cite{Ducourant14} core stars have the oldest age at $3.8\pm1.1\,$Myr. See Section~4 for more details.}
	\label{fig:subsample}
\end{center}
\end{figure}

The ages we get for the traceback analysis are small for any subsample that we use. The radius of the cluster also does not get much smaller than it currently is. Figure~\ref{fig:subsample} shows the results of several subsamples we used. We compare the radius of TWA as a function of time with three subsamples, 1) All 38 stars with full kinematics, 2) Only those with distance less than 85\,pc (see below for the justification of 85\,pc), and 3) The sample of core stars identified by \cite{Ducourant14} (except TWA 3).  The stars with distances less than 85\,pc diverge the fastest with a minimum size occurring at only $0.8\pm0.3\,$Myr.  The core stars of \cite{Ducourant14} diverge more slowly, but still diverge after only $3.8\pm1.1\,$Myr.

We also considered the jackknife resampling technique used by \cite{Ducourant14}.  We reran our traceback code 10 times, randomly removing 10 objects from our list of 38 for each run. There were no significant differences between each run, and the average age was 0.1\,Myr, with the oldest being only 0.2\,Myr.  

\begin{figure}[ht!]
\begin{centering}
	\includegraphics[width=0.87\textwidth]{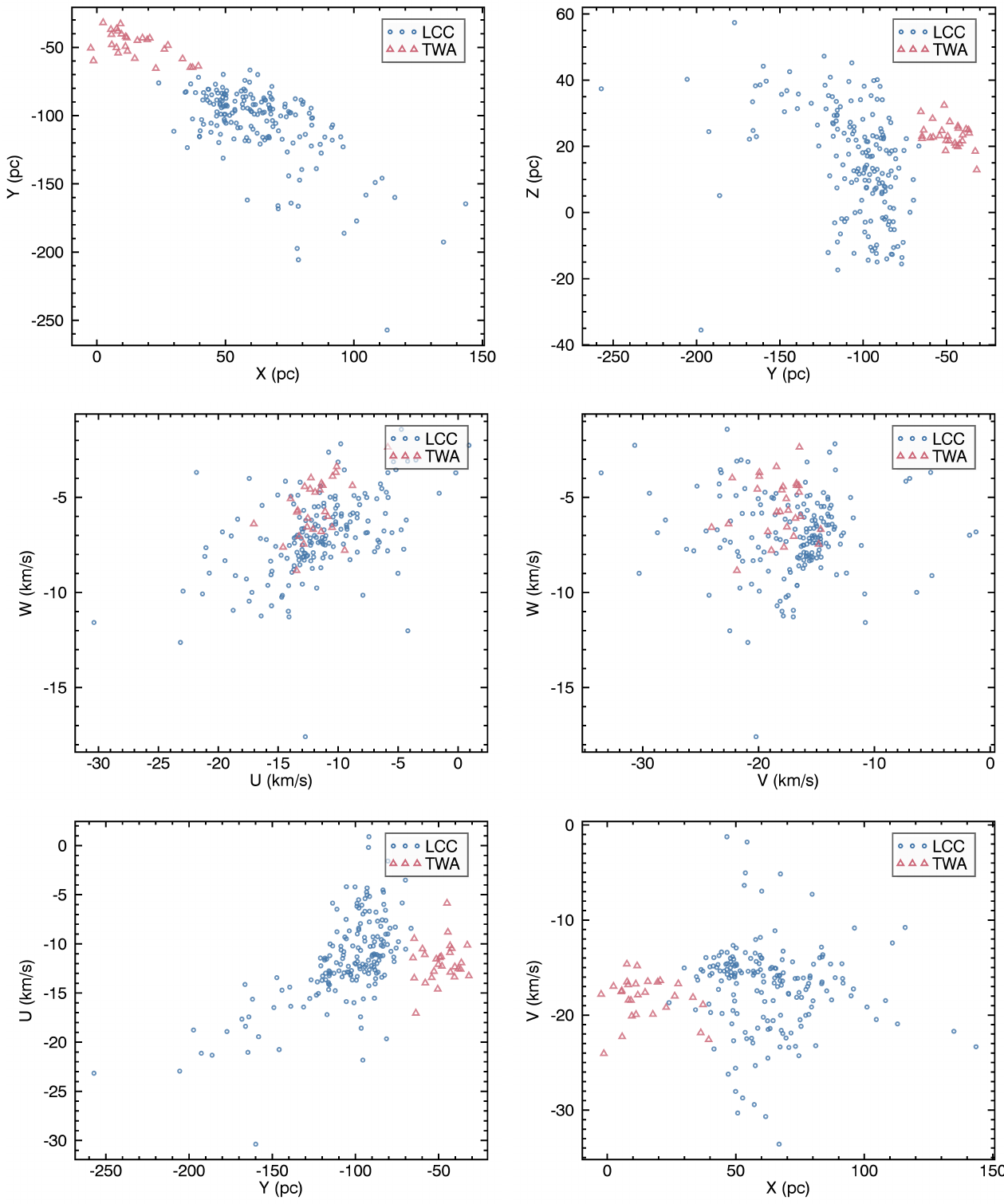}
	\label{fig:corr}
	\caption{Galactic $XYZ$ positions and $UVW$ velocities of LCC (blue circles) and TWA (red triangles). The two associations occupy the same region in velocity space, and are nearby in position.  See Section~4 for more details.}
\end{centering}
\end{figure}

To test the TWA kinematics, we want to exclude stars that are ambiguous members of TWA. 
Several stars in our sample may be members of Lower Centaurus Crux (LCC), which shares many similar characteristics as TWA in $XYZ$ positions and $UVW$ velocities. Some stars in our sample, such as TWA 17 and 18, have distances that would make them more likely to be LCC members than TWA members, which has been known for some time \citep[e.g.][]{Mamajek05}.  Figure~\ref{fig:corr} shows the Galactic positions and velocities of proposed LCC and TWA members.  In velocity space, the two groups are more or less indistinguishable.  In position, the groups are more separated, but some proposed TWA members appear more closely associated with LCC. \cite{Mamajek05} describes the marker between where TWA ends and LCC begins as a distance of 85\,pc.  This criterion would exclude TWA 14, 15A, 15B, 17, 18, 19A and 2M1025--42 from being members of TWA.  

To account for the contamination of other possible non-members, we reran the traceback analysis by tossing out incompatible stars.  We tossed out stars one at a time by removing the star that is farthest away from the mean cluster position at the time when the stars are the closest.  The first stars to be tossed out are TWA 18, 17, 15B 15A, 2M1025--42, and 14, in that order.  At that point, the traceback age for the remaining stars is 0.7\,Myr.  After that, the traceback age gets progressively smaller as more and more stars are tossed out.  

\subsection{Effect of measurement uncertainties on the traceback analysis\label{sub:MC}}

When we did our traceback analysis on the group of ``core stars'' identified by \cite{Ducourant14}, we got an age of $3.8\pm1.1\,$Myr.  However, \cite{Ducourant14} get an age of $7.5\pm0.7$\,Myr when they do a similar analysis.  This stark difference comes from our Monte Carlo method of dealing with the measurement uncertainties.  If we remove the Monte Carlo portion of our analysis, we get an age of 7.9\,Myr for $R_\text{mean}$ and 6.9\,Myr for $R_\text{std}$.  

These 15 core stars were selected by \cite{Ducourant14} to be the ones that best give a convergence, which holds true even with our new parallax measurements.  However, if we remove the most distant stars, TWA 14, 15A, and 15B, i.e.\ the ones most likely to be members of LCC, not TWA, then the age without the Monte Carlo treatment drops to 3.3\,Myr for $R_\text{mean}$ and 3.2\,Myr for $R_\text{std}$.  Using all 38 stars in Table~\ref{tab:velocities}, we get 2.5\,Myr for $R_\text{mean}$ and 1.5\,Myr for $R_\text{std}$.  

\begin{figure}[t!]
\begin{center}
	\includegraphics[width=\textwidth]{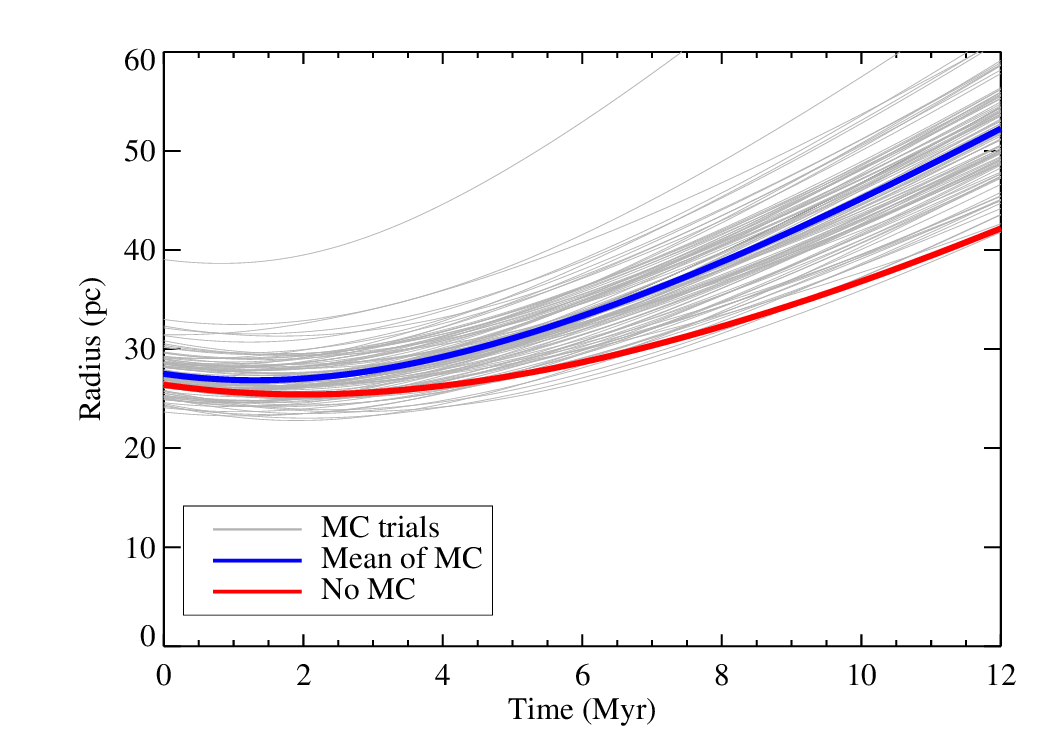}
	\caption{Random selection of 100 out of 10,000 trials of the Monte Carlo traceback in gray with the mean of all 10,000 trials in blue.  The red curve shows the traceback of all 38 TWA stars without using a Monte Carlo method.  The average MC trial is more divergent than the non-MC traceback. See Section~\ref{sub:MC} for more details.}
	\label{fig:100}
\end{center}
\end{figure}

By including the uncertainty in our measurements in our Monte Carlo analysis, we are effectively taking into account a dispersion which makes it harder to accurately trace the stars back.  Figure~\ref{fig:100} shows 100 random trials in black compared to the traceback without the Monte Carlo method in red.  The average trial diverges faster than the non-MC traceback. 

To test how the measured uncertainty affects the accuracy of the traceback, we modeled an idealized version of TWA and attempted to trace it back.  We started with a compact version of TWA, created by halving the distance each star was away from the mean star position.  We then gave every star the $UVW$ velocities of the mean of all 38 TWA candidate members, given in Table~\ref{tab:meanUVW}.  We added an initial velocity dispersion, $\sigma_\text{init}$, and calculated the positions of the stars after 10\,Myr.  We then simulated our measurement uncertainties by including a random component to the velocities taken from a normal distribution with a standard deviation equal to the $\sigma_\text{meas}$.  Using the new stellar velocities, we traced the stars back to calculate a convergence time. 

\begin{figure}[t!]
\begin{center}
	\includegraphics[width=\textwidth]{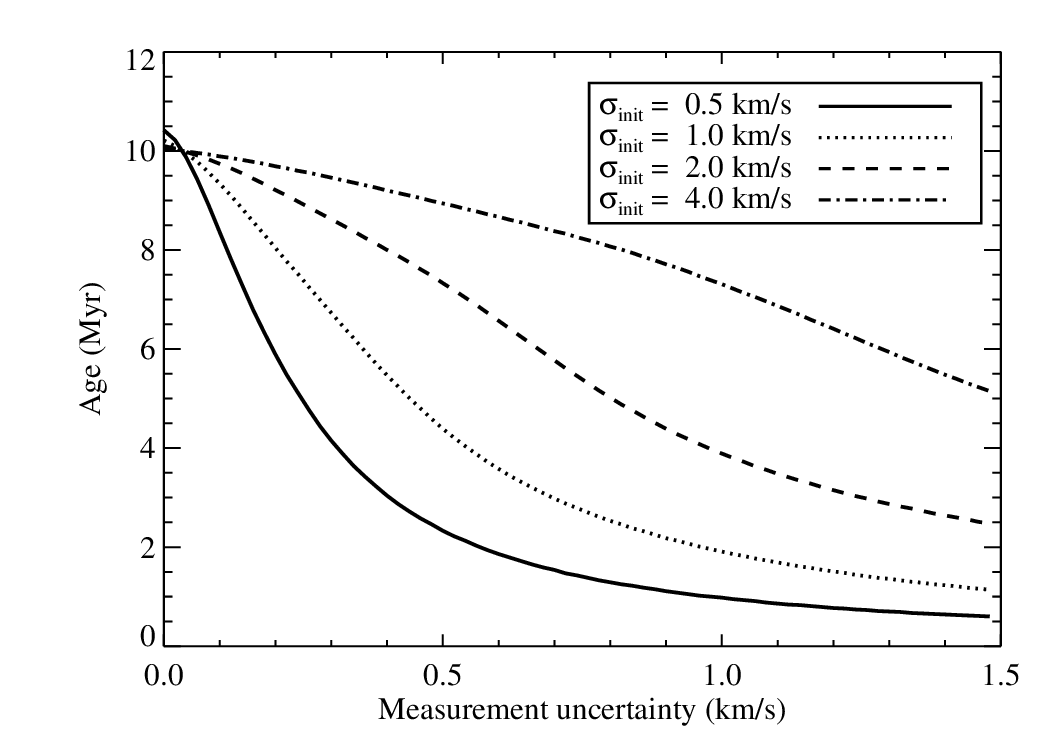}
	\caption{Simulation of a theoretical 10\,Myr-old TWA-like association with four different initial velocity dispersions, $\sigma_\text{init}$.  The curves show the traceback age that is derived after randomly adding noise that simulates measurement uncertainties.  The derived traceback age drops steeply with increasing measured uncertainties. The total measurement uncertainty in our data is dominated by uncertainties in the RV and parallax measurements. See Section~\ref{sub:MC} for more details.}
	\label{fig:mintime}
\end{center}
\end{figure}

Figure~\ref{fig:mintime} shows the calculated traceback time as a function of the measurement uncertainties, $\sigma_\text{meas}$, for four different initial velocity dispersions.  The results show a quick drop-off in the calculated traceback age with $\sigma_\text{meas}$.  As $\sigma_\text{init}$ decreases, the drop-off is steeper, and thus more precisely measured velocities are required to find the true convergence time.

Typical expected velocity dispersions for open clusters are around 1--2\,km\,s$^{-1}$ \citep{Adams00}. Assuming an initial velocity dispersion for TWA of $\sigma_\text{init} = 2$\,km\,s$^{-1}$, we would need to measure the $UVW$ velocities with a precision of 0.25\,km\,s$^{-1}$ to accurately measure the traceback time to within 1\,Myr.  Given our current mean measurement uncertainties of $\sigma_\text{meas} = 1.3$\,km\,s$^{-1}$, the initial velocity dispersion of TWA would have to be greater than 11\,km\,s$^{-1}$ for us to accurately measure the traceback age to within 1\,Myr.  

The GAIA spacecraft will soon obtain parallaxes to a billion stars with unprecedented precision.  Given the low uncertainties in GAIA parallaxes and proper motions, the RVs will become the dominate source of noise.  From the calculations above, the RVs will need to have a precision of 0.25\,km\,s$^{-1}$ for a proper traceback analysis, a quarter of our current average RV uncertainty.  

We ran another traceback analysis, this time with only stars with RV uncertainties below 0.5\,km\,s$^{-1}$ (17 stars).  This produces a traceback age of $2.9\pm0.5\,$Myr. This is a slightly higher value than most of the subsamples tested, but still falls quite short of the expected value.  Given the low traceback ages that we obtain for TWA, it is likely that its initial velocity dispersion is too low for an accurate determination of its age using a traceback analysis.

\section{Pre-main sequence track age}

\subsection{Calculating the ages of individual stars\label{sub:ages}}

To determine the ages of individual TWA stars, we compare the stars to the pre-main sequence tracks of \cite{Baraffe15}. We use our parallax values to convert 2MASS apparent magnitudes to absolute magnitudes.  We took optical spectral types from the literature and converted them to effective temperature using Table~6 of  \cite{Pecaut13}.  For stars M5--M7, we use the main-sequence values in Table~5 of \cite{Pecaut13}. It is interesting to note the disagreement between \cite{Pecaut13} and \cite{Luhman99}, which was used in \cite{Weinberger13} to convert to effective temperature.  The disagreement is typically about 100\,K cooler for \cite{Pecaut13} than for \cite{Luhman99}, with particular disagreement around M5 with a difference of about 250\,K.  For stars with spectral type greater than M7, we used the polynomial relation in Table~19 of \cite{Faherty16} for moving group members.  
 Spectral types and temperatures for the entire sample are listed in Table~\ref{tab:ages}. 

The uncertainties on the temperature are derived from the spread in temperatures from the uncertainties in the spectral type, as given in the references we cite. Some of the uncertainties in temperature derived this way are much larger than the typically assumed values of 75--100\,K. 

For each star, we then use the absolute magnitudes in $J$, $H$, and $K$, and the temperature to derive an age by interpolating the \cite{Baraffe15} models.  The uncertainties on the ages are calculated from the uncertainties on both the parallaxes and the temperature. 

\begin{figure}[t!]
\begin{center}
	\includegraphics[width=0.9\textwidth]{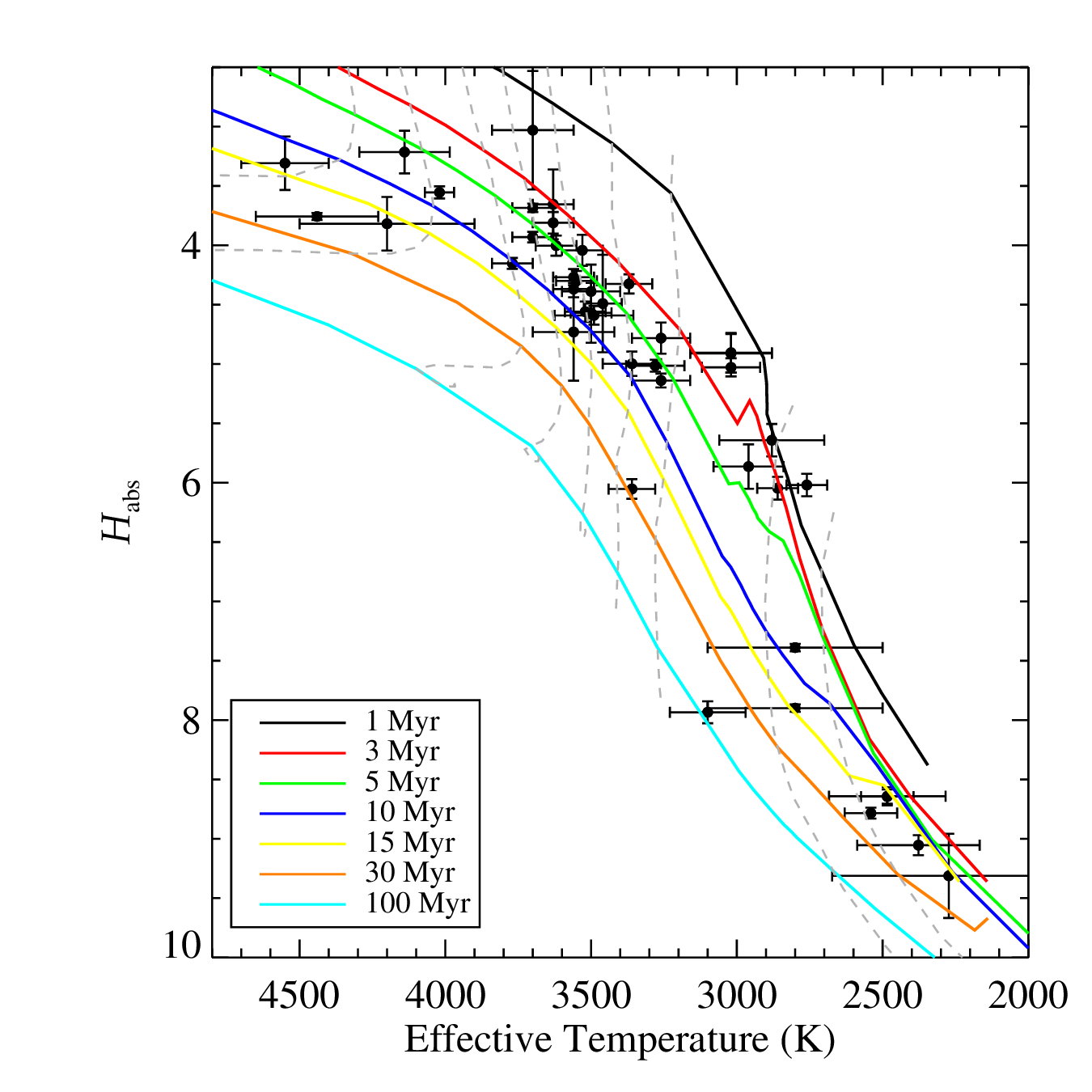}
	\caption{Absolute $H$ magnitude vs. effective temperature of the TWA stars plotted against the \cite{Baraffe15} pre-main sequence evolutionary tracks. Most of the stars are clustered between 3 and 10\,Myr. See Section~\ref{sub:ages} for more details.}
	\label{fig:Hmag}
\end{center}
\end{figure}

\begin{figure}[t!]
\begin{center}
	\includegraphics[width=\textwidth]{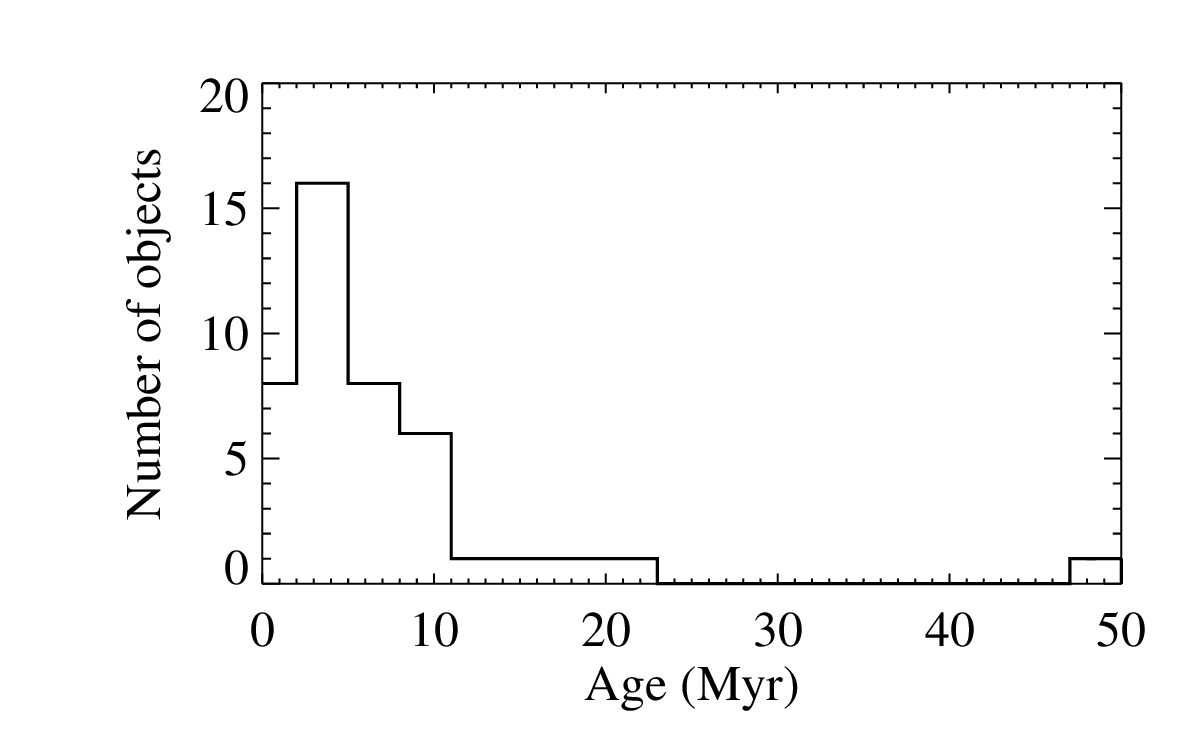}
	\caption{Histogram of the ages of TWA members as given in Table~\ref{tab:ages}. The bin size is 3\,Myr.  The majority of stars in TWA have ages $\lesssim$ 10\,Myr, but there is a tail that extends past 20\,Myr.  There is one clear outlier, TWA 31, which has an age of 50\,Myr. See Section~\ref{sub:ages} and Table~\ref{tab:ages} for more details.}
	\label{fig:histo_ages}
\end{center}
\end{figure}

Figure~\ref{fig:Hmag} shows the absolute $H$ magnitude vs.\ temperature of all the stars in our sample with the \cite{Baraffe15} models overplotted. The interpolated ages are given in Table~\ref{tab:ages}, along with their uncertainties. 
The histogram of ages is shown is Figure~\ref{fig:histo_ages}. 

An Anderson-Darling test of the histogram of ages confirms that the distribution of ages is not normal. We determine the mean age and its uncertainty by bootstrapping. The sample has a mean age of $7.9\pm1.0$\,Myr and a median age of 6.2\,Myr.

We also compared our data to the older \cite{Baraffe98} models to see if there have been any major changes between models. We find them to be in good agreement with the \cite{Baraffe15} models, and we get a mean age of $7.8\pm1.0\,$Myr for all stars. 

There is a clear outlier in Figure~\ref{fig:histo_ages}, TWA 31, which has an age of 50\,Myr. TWA 31 is not a bona fide member of TWA, and has only recently been proposed as a member of TWA \citep{Shkolnik11}.  It  may be a member of the older association LCC (J.~Gagn{\'e} et al.\ submitted to ApJS). 

Another star of note is TWA 9A.  Its parallax values from GAIA and Hipparcos differ by more than 6\,mas, which drastically changes its absolute magnitude, and hence its derived age. With the new GAIA parallax, TWA 9A has an age of $7\pm4\,$Myr, typical for a TWA member. But with the old Hipparcos value, its age is 30\,Myr, which would make it an outlier.  GAIA is more precise than Hipparcos, but it is unclear why the parallax values for this particular star are so discrepant.  Surprisingly, the \cite{Ducourant14} parallax value is consistent with the Hipparcos value and more than 2$\,\sigma$ higher than the GAIA value.

\subsection{Spread in the ages of TWA members}
Even if outliers are excluded, a spread is observed in the ages of TWA stars of a few Myr (see Fig.\ 8). This could be an intrinsic spread in stellar ages, or it could be due to uncertainties in the calculations due to several factors.

First, there is the uncertainty in the spectral typing, which propagates through to the age via the temperature.  Some of these uncertainties are quite large ($\sim200$--300\,K), making the location on the pre-main sequence track highly uncertain. The spectral types were also not all taken from the same sample. For consistency, we took the spectral types from \cite{Herczeg14} where available (more than a third of our sample), but we had to search the literature to find the rest. 

The next uncertainty is the conversion from spectral type to effective temperature.  We use the tabulated values of \cite{Pecaut13}.  Other models are available that will give different values.  The estimated temperature uncertainty between models is 100\,K. \cite{Weinberger13} did a similar study of a subsample of the TWA targets, and their ages for the same stars tend to be older. They used a different conversion from spectral type to effective temperature, \cite{Baraffe98} tracks, and different spectral types for some of the stars. We use their parallaxes, but our temperatures are systematically cooler, making our resulting ages younger.  \cite{Weinberger13} also get a wider spread in ages with an older mean age.

As mentioned previously, there is a disagreement between the methods of \cite{Pecaut13} and \cite{Luhman99} for converting spectral types to temperature for stars M5 and later.  We have thus looked at a subsample consisting only of stars with spectral types $\leq$ M4 to see if there is a better agreement in the estimated ages of TWA members.  We found an average age of $7.5\pm0.7\,$Myr with a median age of 6.2\,Myr.  

The picture is even more complicated when looking at the later spectral types. Sources on the M/L boundary tend to show a sharp change in physical properties when moving to later spectral types \citep{Faherty16}. For these low temperature objects, we are likely probing the region where clouds start to dominate. These objects should be treated differently than the higher mass stars.  

To determine if the spread in ages is due to an intrinsic spread or due to measurement uncertainties, we ran a Monte Carlo simulation of the association. We created 10,000 synthetics sets of stars, all with an age of 10\,Myr.  We then added a random noise component to each star, taken from a normal distribution centered around zero with the age uncertainty of each star as the standard deviation of the noise distribution.  The measured spread is then the average of the spreads of each trial.  An association with no intrinsic age spread would still have a measured distribution with a standard deviation of $2.7\pm1.0$\,Myr given our uncertainty in the measurement. This is consistent with the spread seen in Fig.\ 8.

In addition to the observational uncertainties, there are systematic uncertainties involved in modeling of pre-main sequence stars that can cause the spread in ages. First is the accretion history of the stars.  Stocastic accretion events early in a star's life can produce luminosity changes that can explain the spread in the color-magnitude diagrams of young stellar associations \cite{Baraffe12}.  This appears as an age spread even when none exists.

Various other factors also affect the modeling of pre-main sequence stars, including effects of changing surface gravity as the stars move onto the main sequence, the presence of disks and their impact on accretion and extinction, and stellar activity such as star spots \citep{Soderblom14}. Different stellar evolutionary models will produce different isochrones, a result that is much more pronounced for young, low mass stars, which describes much of our sample \citep{Hillenbrand08}. Additional factors such as transforming models to the observational plane and extinction corrections can also produce an observed age spread. At this point, it is not possible to conclude if there is an intrinsic age spread or not.

\section{Discussion\label{sub:discussion}}
The best way to determine if a candidate member of an association is a true member is to compare its properties to those of the core members. Given the inability to find a convergence with any subset of stars, there is no good way to determine core members that is not somewhat arbitrary. We take the approach of starting with the full 38 candidates we have, and cut them down by applying thresholds.

The first step is to cut out the stars that are more likely associated with the nearby LCC.  \cite{Mamajek05} defines the border with LCC to be at a distance of 85\,pc, and Gagn{\'e}~et~al.~(submitted to ApJS) refines this to 80\,pc.  So we start by excluding the most distant stars, TWA 14, 15A, 15B, 17, 18, 19A, 31, and 2M1025--42.  Next, to ensure accuracy in our TWA core sample, we remove all stars with RV uncertainties great than 1.0\,km\,s$^{-1}$.  This criterion excludes TWA 5, 6, 20, 26, 28, 29, 30A, and 43.  

Uncertainty in parallax also affects the accuracy of the traceback, so we also eliminated sources with parallax uncertainties greater than 1.0\,mas. This removes many stars from our core group that are usually considered bona fide members.  We are not implying these stars are not members of TWA, we are simply trying to choose a core group that has the most precise measurements and whose members we are sure are related.

Lastly, we calculate a mean $UVW$ for the remaining stars and remove the sources with velocities more than 10\,$\sigma$ from the mean in any direction.  This leaves 14 stars that we define as the core members, TWA 1, 8A, 8B, 9A, 10, 11A, 12, 13A, 13B, 16, 23, 25, 27, 32.  The mean $UVW$ velocities of these subsamples at each cut and their mean ages are listed in Table~\ref{tab:meanUVW}.  

These core members can be used to constrain models that predict membership probabilities.  Using the current BANYAN II tool \citep{Gagne14, Malo13}, we calculate the probabilities of membership of our candidate members from \cite{Rodriguez11} and \cite{Riaz06}.  We find that two have high probability of TWA membership, (2M1159-45: 99.7\% probability; 2M1235-41: 96.7\% probability) and one has a low probability of membership (2M1311-42: 0.18\% probability).  The last candidate, 2M1025-42, we already ruled out as a TWA member because of its distance, also gets a 0\% probability of membership in BANYAN II.  Based on these results, we assign TWA names to these two new members. TWA numbers 39-44 are defined in Gagn{\'e}~et~al.~(submitted to ApJS), so we assign the names TWA 45 for 2MASS J11592786--4510192, and TWA 46 for 2MASS J12354615--4115531.  It should also be noted that our parallax measurement of TWA 32 (given in Table~\ref{tab:para}) indicates it should be considered a bona fide member of TWA.

\section{Summary}
We measured the parallaxes of 14 TWA candidate stars, bringing the total number of TWA candidates with measured parallaxes to 48. Of these, 38 have proper motion and RV measurements as well, allowing for a full characterization of their kinematics.  TWA stars are easily confused with the nearby, young association, LCC. We still have no firm way of distinguishing the two groups that is not somewhat arbitrary.  

We have shown that a traceback analysis of TWA does not produce a convergence.  Our Monte Carlo analysis suggests the uncertainties in the parallax and RV measurements are currently too large to accurately traceback the stars.  Furthermore, if TWA had a low initial velocity dispersion ($\lesssim2$\,km\,s$^{-1}$), then the RV measurements would need to be more precise than 0.25\,km\,s$^{-1}$ for an accurate traceback of TWA.

\vspace{1cm}We would like to thank the staff of the Las Campanas Observatory and especially the operators of the du Pont telescope. This work was partially supported by the National Science Foundation under Grant Number 1313043 and by the NASA Astrobiology Institute under cooperative agreement NNA09DA81A. This work makes use of the Simbad database and Vizier catalogue access tool, CDS, Strasbourg, France, and the 2MASS survey, which is a joint project of the University of Massachusetts and the Infrared Processing and Analysis Center/California Institute of Technology, funded by the National Aeronautics and Space Administration and the National Science Foundation. This work has made use of data from the European Space Agency (ESA)
mission {\it Gaia} (\url{http://www.cosmos.esa.int/gaia}), processed by
the {\it Gaia} Data Processing and Analysis Consortium (DPAC,
\url{http://www.cosmos.esa.int/web/gaia/dpac/consortium}). Funding
for the DPAC has been provided by national institutions, in particular
the institutions participating in the {\it Gaia} Multilateral Agreement.

\begin{deluxetable}{lccl}
\tablewidth{0pt}
\tablecaption{Record of Observations}
\tabletypesize{\scriptsize}
\tablehead{
\colhead{Designation} &\multicolumn{2}{l}{Integration Times} & \colhead{Epochs} \\
 \cline{2-3}	& \colhead{FF\tablenotemark{a}} & \colhead{GW\tablenotemark{b}} \\
			& \colhead{(s)} & \colhead{(s)} & \colhead{(JD)} \\
}
\startdata
TWA 6 	& 20	& 2	& 2455635.6, 2456090.5,	2456407.6, 2456429.4, 2456704.6, 2456771.5, 2457039.8, 2457081.6, 2457178.5\\
TWA 8	& 60 & 0.8 & 2456089.5, 2456357.8, 2456428.6, 2456492.4, 2456705.8, 2457080.7, 2457177.5	\\
TWA 10	& 60 & 0.5 & 2455635.8, 2456086.6, 2456357.8, 2456428.6, 2456490.5,  2456704.8, 2457176.5	\\
TWA 17	& 40 & 2	& 2456088.3, 2456492.5, 2456704.9, 2456769.7, 2456805.5,  2456852.5, 2457176.5	\\
TWA 18	& 60 & 0.5 & 2455635.8, 2456086.6, 2456428.6, 2456490.5, 2456704.9, 2456705.9, 2456769.7, 2456852.5, 2457176.6 \\
TWA 30 A	& 40 & 10 & 2455635.7, 2456088.2, 2456428.6, 2457080.7, 2457176.5, 2457472.7	\\
TWA 31	& 40 & 20 & 2455404.5, 2455634.7, 2455663.6, 2455941.8,  2456086.5, 2457176.5, 2457472.8 	\\
TWA 32	& 54 & 6	& 2455634.8, 2455663.7, 2455941.9, 2456086.5, 2456358.8, 2457080.8, 2457176.5 	\\
TWA 33	& 60	& 1	& 2456772.6, 2456852.5, 2457080.7, 2457175.5, 2457410.8 	\\
2M1025--42	& 21 & 0.3 & 2456771.5, 2457081.6, 2457177.5, 2457374.9, 2457443.7, 2457471.5  	\\
2M1159--45	& 40	& 1	&  2456772.5, 2456851.5, 2457080.7, 2457175.5, 2457410.8, 2457443.8, 2457471.6  	\\
2M1235--41	& 50	& 1	&  2456771.7, 2456805.6, 2456851.5, 2457080.8, 2457175.5, 2457410.8, 2457471.7 	\\
2M1311--42	& 48 & 1.5 & 2456771.7, 2456805.6, 2456851.5, 2457080.8, 2457410.9 2457471.8 	\\
\enddata
\label{tab:obs}
\tablenotetext{a}{FF: Full Field}
\tablenotetext{b}{GW: Guide Window}
\tablecomments{See Section~2 for more details}
\end{deluxetable}

\begin{deluxetable}{lrrrrrr}
\tablewidth{0pt}
\tabletypesize{\footnotesize}
\tablecaption{Measured Parallaxes and Proper Motions}
\tablehead{
\colhead{Designation} & \colhead{$\pi_\text{rel}$}  & \colhead{Zero Point} & \colhead{$\pi_\text{abs}$} & \colhead{Distance} 
& \colhead{$\mu_\text{ra}$} & \colhead{$\mu_\text{dec}$} \\
			 & \colhead{(mas)}  & \colhead{(mas)} & \colhead{(mas)} & \colhead{(pc)} &
			  \colhead{(mas\,yr$^{-1}$)} & \colhead{(mas\,yr$^{-1}$)} 
}
\startdata
TWA 6	&$	16.36\pm  0.97 	$ &$	 -0.42\pm     	0.32$&	$16.78\pm 1.02$ &	$		59.59\pm      3.6$  & $-49.81 \pm1.09$  & $-18.56\pm0.91$\\	
TWA 8 A	&$	20.90\pm     1.19$&$	 -0.31\pm     	0.38$&	$21.21\pm 1.25$ &	$		47.15\pm      2.8$ & $-81.19\pm2.06$ &$ -27.61\pm 4.53$\\
TWA 8 B	&$	21.60\pm      2.38$&$ -0.08\pm    	0.36$&	$21.68\pm 2.41$ &	$		46.123\pm    5.1$ & $-79.02\pm31.3$ & $ -23.36 \pm 8.36$ \\
TWA 10	&$	16.53\pm     0.72$&$	 -0.41\pm     	0.15$&	$16.94\pm 0.73$ & 	$		59.03\pm      2.5$ & $-58.90\pm0.54$ & $-29.58\pm 0.94$\\
TWA 17	&$	6.02\pm     0.50$&$	 -0.37\pm     	0.18$&	$6.39\pm 	0.53$ &	$		156.50\pm  13.0$ & $-22.04\pm1.19$ & $-18.42 \pm2.62$ \\
TWA 18	&$	6.09\pm      1.41$&$	 -0.09\pm	 	0.10$&	$6.18\pm	1.42$&	$		161.87\pm   37.2$& $-24.04\pm0.72$ & $-20.42 \pm1.70$ \\
TWA 30 A	&$	20.55\pm      1.31$&$ -0.46\pm   	0.16$&	$21.01\pm 1.32$&	$		47.60\pm      3.0$ & $-83.19\pm0.76$ & $-24.42\pm1.01$\\
TWA 31	&$	12.40\pm     0.42$&$  0.13\pm      	0.31$&      $12.27\pm0.52$&     $ 		81.48\pm      3.5$ & $-38.92\pm0.16$ & $-21.63\pm0.40$\\
TWA 32	&$	14.89\pm     0.65$&$	 -0.22\pm	 	0.12$&	$15.11\pm0.66$&	$		66.14\pm      2.9$ & $-55.34\pm0.62$ & $-27.97\pm2.81$\\
TWA 33	&$	19.44\pm     1.58$&$	 -0.06\pm 		0.55$&	$19.50\pm1.68$&	$		51.28\pm      4.4$ & $-79.00\pm6.29$ & $-21.78\pm3.95$\\
2M1025--42&$	9.85\pm      1.22$&$ 	0.55\pm            0.29$& 	$9.30\pm	1.26$&      $		107.50\pm    14.5$& $-34.78\pm2.15$ & $-3.08\pm1.41$\\
2M1159--45&$	13.29\pm     0.93$&$ 0.36\pm          0.35$&	$12.93\pm0.99$&	$		77.34\pm      5.9$ & $-46.67\pm6.28$ & $-18.88\pm2.07$\\
2M1235--41&$	20.27\pm     0.77$&$	-0.48\pm         0.19$&	$20.75\pm0.79$&	$	 	48.19\pm      1.8$ & $-58.61\pm1.64$ & $-24.82\pm2.15$\\
2M1311--42&$	12.19\pm     0.93$&$	-0.34\pm        0.17$&		$12.53\pm0.94$&	$		79.80\pm      6.0$ & $-29.62\pm0.87$ & $-19.91\pm0.68$\\
\enddata
\label{tab:para}
\tablecomments{See Section~3 for more details}
\end{deluxetable}

\begin{deluxetable}{lrrrrrrrrr}
\tablewidth{0pt}
\tablecaption{Positions and Velocities Used in Traceback Analysis}
\tabletypesize{\tiny}
\tablehead{
\colhead{Designation} & \colhead{R.A.}  & \colhead{Decl.} & \colhead{$\pi_\text{abs}$} & \colhead{Ref} & \colhead{RV} & \colhead{Ref}
		& \colhead{$\mu_\text{ra}$} & \colhead{$\mu_\text{dec}$} & \colhead{Ref} \\
	& \colhead{(h m s)} & \colhead{(d m s)} & \colhead{(mas)} &  & \colhead{(km\,s$^{-1}$)} & & \colhead{(mas\,yr$^{-1}$)} &
		\colhead{(mas\,yr$^{-1}$)} &
			 }
\startdata
TWA 1 & 11 01 51.91 & -34 42 17.03 & $16.8\pm0.3$ & 1  & $12.66\pm0.22$ & 10 & $-69.27\pm0.870$ & $-11.58\pm 1.03$ & 2,26,27  \\
TWA 2 & 11 09 13.80 & -30 01 39.88 & $21.6\pm1.3$ & 2,3  & $10.98\pm0.03$ & 11 & $-88.38\pm0.740$ & $-19.19\pm0.890$ & 2,26,27  \\
TWA 4 & 11 22 05.29 & -24 46 39.76 & $22.3\pm2.3$ & 4  & $9.25\pm0.69$ & 10,12,13  & $-91.43\pm 1.52$ & $-31.75\pm 1.05$ & 26,27  \\
TWA 5 & 11 31 55.26 & -34 36 27.24 & $20.7\pm0.7$ & 1  & $13.3\pm2.0$ & 14 & $-78.15\pm0.650$ & $-21.46\pm0.600$ & 26,27  \\
TWA 6 & 10 18 28.70 & -31 50 02.85 & $15.7\pm0.3$ & 1  & $20.39\pm1.44$ & 10,15 & $-55.20\pm0.950$ & $-19.92\pm 1.10$ & 26,27  \\
TWA 7 & 10 42 30.11 & -33 40 16.21 & $29.0\pm2.1$ & 2  & $12.13\pm0.20$ & 11,14 & $-115.2\pm0.750$ & $-20.19\pm0.750$ & 2,26,27  \\
TWA 8 A & 11 32 41.26 & -26 51 55.99 & $21.5\pm0.8$ & 2,5,6  & $8.34\pm0.48$ & 14 & $-87.10\pm0.800$ & $-28.00\pm0.800$ & 2  \\
TWA 8 B & 11 32 41.17 & -26 52 09.13 & $22.6\pm1.0$ & 2,5,6  & $8.93\pm0.27$ & 14 & $-86.50\pm0.900$ & $-25.00\pm0.900$ & 2  \\
TWA 9 A & 11 48 24.23 & -37 28 49.11 & $13.2\pm0.3$ & 1  & $10.39\pm0.62$ &11  & $-54.81\pm 1.98$ & $-19.63\pm 1.97$ & 2,26  \\
TWA 9 B & 11 48 23.73 & -37 28 48.50 & $19.2\pm1.1$ & 2  & $11.51\pm0.89$ & 11 & $-51.00\pm0.600$ & $-18.10\pm0.600$ & 2  \\
TWA 10 & 12 35 04.25 & -41 36 38.64 & $16.7\pm0.6$ & 2,5  & $6.75\pm0.40$ & 14 & $-65.24\pm0.380$ & $-21.30\pm0.380$ & 2,27  \\
TWA 11 A & 12 36 01.03 & -39 52 10.23 & $13.7\pm0.3$ & 4  & $7.212\pm0.70$ & 10,13,16 & $-55.44\pm0.790$ & $-23.28\pm0.740$ & 26,27  \\
TWA 12 & 11 21 05.48 & -38 45 16.51 & $15.6\pm0.6$ & 2,3  & $10.94\pm0.32$ & 11,14 & $-66.24\pm0.470$ & $-7.830\pm0.470$ & 2,27  \\
TWA 13 A & 11 21 17.22 & -34 46 45.47 & $18.0\pm0.7$ & 3  & $10.8\pm0.4$ & 11 & $-65.70\pm 4.35$ & $-12.80\pm 4.10$ & 3  \\
TWA 13 B & 11 21 17.45 & -34 46 49.83 & $16.8\pm0.7$ &3   & $11.48\pm0.4$ & 11 & $-67.30\pm 4.77$ & $-11.40\pm 4.52$ & 3  \\
TWA 14 & 11 13 26.22 & -45 23 42.75 & $10.4\pm1.2$ & 3  & $15.83\pm2.00$ & 14  & $-43.90\pm 1.40$ & $-7.400\pm 1.40$ & 27  \\
TWA 15 A & 12 34 20.65 & -48 15 13.48 & $9.1\pm1.7$ & 3  & $11.2\pm2.0$ & 15 & $-36.80\pm 4.31$ & $-10.70\pm 4.18$ & 3  \\
TWA 15 B & 12 34 20.47 & -48 15 19.59 & $8.6\pm1.6$ & 3  & $10.03\pm1.66$ & 14 & $-35.80\pm 4.61$ & $-10.20\pm 4.61$ & 3  \\
TWA 16 & 12 34 56.30 & -45 38 07.63 & $12.8\pm0.5$ & 3  & $8.85\pm0.27$ & 11,14 & $-49.40\pm 4.35$ & $-26.10\pm 5.87$ & 3  \\
TWA 17 & 13 20 45.39 & -46 11 37.7 & $6.4\pm0.5$ & 5  & $8.09\pm3.46$ & 15,17 & $-31.30\pm 1.00$ & $-17.70\pm 1.10$ & 27  \\
TWA 18 & 13 21 37.23 & -44 21 51.85 & $6.2\pm1.4$ & 5  & $4.32\pm2.27$ &15,17  & $-32.10\pm 1.10$ & $-20.40\pm 1.10$ & 27  \\
TWA 19 A & 11 47 24.55 & -49 53 03.01 & $9.1\pm0.3$ &  1 & $11.95\pm0.54$ & 10,13,18 & $-33.01\pm0.990$ & $-8.120\pm0.990$ & 26,27  \\
TWA 20 & 12 31 38.07 & -45 58 59.47 & $12.9\pm0.6$ &  3 & $9.38\pm3.71$ & 15,17 & $-63.50\pm 1.10$ & $-27.80\pm 1.10$ & 27 \\
TWA 21 & 10 13 14.77 & -52 30 53.95 & $19.1\pm0.3$ & 1  & $17.76\pm0.23$ & 11,18,19 & $-61.14\pm0.470$ & $ 11.24\pm0.530$ & 2,26,27 \\
TWA 22 & 10 17 26.91 & -53 54 26.42 & $57.0\pm0.7$ & 7  & $13.50\pm0.09$ &11,14  & $-175.8\pm0.780$ & $-20.69\pm0.780$ & 6,27  \\
TWA 23 & 12 07 27.38 & -32 47 00.25 & $18.7\pm0.5$ & 2,3  & $10.91\pm0.10$ & 11,14 & $-73.63\pm0.700$ & $-26.87\pm0.760$ & 2,27  \\
TWA 25 & 12 15 30.72 & -39 48 42.59 & $19.3\pm0.4$ &  1 & $7.03\pm0.58$ & 11 & $-73.52\pm0.750$ & $-27.55\pm0.760$ & 26,27  \\
TWA 26 & 11 39 51.14 & -31 59 21.50 & $25.8\pm1.0$ & 2,3  & $11.6\pm2.0$ & 20 & $-93.30\pm0.500$ & $-27.50\pm0.500$ & 2  \\
TWA 27 & 12 07 33.47 & -39 32 54.00 & $19.0\pm0.4$ & 2,8,9  & $9.00\pm0.96$ & 15,20,21 & $-64.20\pm0.400$ & $-22.60\pm0.400$ & 2  \\
TWA 28 & 11 02 09.83 & -34 30 35.5 & $18.1\pm0.5$ & 7  & $9.0\pm3.3$ & 22 & $-67.24\pm0.600$ & $-14.02\pm0.600$ & 7,28  \\
TWA 29 & 12 45 14.16 & -44 29 07.7 & $12.7\pm2.1$ & 3  & $7.7\pm3.2$ & 22 & $-71.00\pm 7.00$ & $-23.00\pm 3.00$ & 2,3  \\
TWA 30 A & 11 32 18.31 & -30 19 51.85 & $21.0\pm1.3$ & 5  & $12.3\pm1.5$ & 23 & $-87.80\pm 1.30$ & $-25.20\pm 1.30$ & 27  \\
TWA 31 & 12 07 10.89 & -32 30 53.72 & $12.3\pm0.5$ & 5  & $10.47\pm0.41$ &14  & $-46.92\pm 4.00$ & $-20.83\pm 4.02$ & 5  \\
TWA 32 & 12 26 51.37 & -33 16 12.55 & $15.1\pm0.7$ & 5  & $7.15\pm0.26$ & 14 & $-61.60\pm 4.90$ & $-23.00\pm 2.00$ & 27  \\
TWA 33 & 11 39 33.83 & -30 40 00.29 & $19.5\pm1.7$ & 5 & $5.8 \pm0.7 $ & 22 & $-79.0\pm 5.1$ & $-25.2\pm 1.8$ & 27 \\
TWA 39 A & 10 12 09.08 & -31 24 45.1 & $18.5 \pm1.7$ & 6 & $14.7 \pm0.5$ & 24 & $-74.8\pm1.1$ & $-9.4\pm1.0$ & 27 \\
TWA 43 & 11 08 44.00 & -28 04 50.4 & $18.0\pm0.5$ & 4 & $16.0 \pm7.4$ & 25 & $-73.3\pm0.6$ & $-21.4\pm0.8$ & 26,27 \\
2M1025-42 & 10 25 20.92 & -42 41 54.0 & $9.3\pm1.3$ & 5  & $17.6\pm0.5$ & 24 & $-45.80\pm 1.20$ & $-2.500\pm 1.20$ & 27  \\
\enddata
\label{tab:velocities}
\tablerefs{(1)~GAIA Data Release 1 \citep{GAIAdr1}, (2)~\cite{Ducourant08,Ducourant14}, (3)~\cite{Weinberger13}, (4)~Hipparcos \citep{Perryman97} 
(5)~This work,
	(6)~\cite{Riedel14}, (7)~\cite{Teixeira08,Teixeira09}, (8)~\cite{Biller07}, (9)~\cite{Gizis07}, 
	(10)~\cite{Torres03,Torres06}, (11)~\cite{Elliott14}, (12)~\cite{Karatas04}, (13)~\cite{Bobylev07}, (14)~\cite{Shkolnik11},
	(15)~\cite{Reid03}, (16)~\cite{Gontcharov06}, (17)~\cite{Kordopatis13}, (18)~\cite{Song03,Song12}, (19)~\cite{Desidera15},
	(20)~\cite{Mohanty03}, (21)~\cite{Rice10}, (22)~J.~Gagn{\'e} et al.\ (submitted to ApJS), (23)~\cite{Looper10}, (24)~\cite{Malo14},
	(25)~\cite{Kharchenko07}, (26)~Tycho-2 \citep{Hog00}, (27)~UCAC4 \citep{Zacharias13}, (28)~PPMXL \citep{Roeser10}}
\tablecomments{See Section~4 for more details}
\end{deluxetable}

\begin{deluxetable}{llllll}
\tablewidth{0pt}
\tabletypesize{\scriptsize}
\tablecaption{Parameters Used for Isochronal Ages}
\tablehead{
\colhead{Designation} & \colhead{Spectral} & \colhead{Reference} & \colhead{$T_\text{eff}$} & \colhead{$H_\text{abs}$} & \colhead{Age} \\
	& \colhead{Type} & & \colhead{(K)} & \colhead{(mag)} & \colhead{(Myr)}
}
\startdata
TWA 1  & M0.5 & \cite{Herczeg14} &$3700\pm70$& $ 3.68\pm0.04$ &$4\pm3$\\
TWA 2 A & M1.7 & \cite{Herczeg14} & $3530\pm70$ & $4.04\pm0.13$ & $4\pm3$\\
TWA 2 B & M3.5 & \cite{Herczeg14} & $3260\pm100$ & $4.78\pm0.13$ & $6\pm3$\\
TWA 4 Aab & K3 & \cite{Laskar09} & $4550\pm150$ & $3.31\pm0.23$ & $13\pm9$\\
TWA 4 Ba & K4.7 & \cite{Laskar09} & $4200\pm300$ & $3.82\pm0.23$ &$16\pm10$ \\
TWA 4 Bb & M2 & \cite{Laskar09} & $3500\pm100$ & $4.39\pm0.23$ & $6\pm3$ \\
TWA 5 Aa	&M1.5 & \cite{Konopacky07} & $3560\pm 70$ &$4.27\pm0.07$  &$5\pm3$\\
TWA 5 Ab &M1.5 & \cite{Konopacky07} & $3560\pm 70$ & $4.37\pm0.07$ &$7\pm4$\\
TWA 5 B & M8.5&\cite{Lowrance99} &$2480\pm200$ & $8.64\pm0.08$ & $6\pm4$ \\
TWA 6 & M0 & \cite{Herczeg14} & $3770\pm70$ & $4.15\pm0.05$ & $10\pm5$ \\
TWA 8 A & M2.9 &\cite{Herczeg14} & $3370\pm80$ & $4.33\pm0.08 $& $4\pm2$ \\
TWA 8 B & M5.2 &\cite{Herczeg14} & $2860\pm70$ & $ 6.05\pm0.10$ & $3\pm1$ \\
TWA 9 A & K6 & \cite{Herczeg14} & $4020\pm50$ & $3.55\pm0.05$ & $7\pm4$\\
TWA 9 B &M3.4 & \cite{Herczeg14} & $3280\pm100$ & $5.01\pm0.05$ &$6\pm3$\\
TWA 10 & M2 & \cite{Torres06} & $3490\pm140$ & $4.59\pm0.08$ & $ 7\pm4$\\
TWA 11 C &M4.5&\cite{Kastner08}&$3020\pm100$ & $5.03\pm0.08$ & $1\pm1$\\
TWA 12 & M1.6 &\cite{Shkolnik11}&$3550\pm70$ & $4.30\pm0.08$ & $6\pm3$\\
TWA 13 A & M1.1& \cite{Herczeg14}&$3620\pm70$ & $4.00\pm0.08$ & $6\pm3$\\
TWA 13 B & M1.0 & \cite{Herczeg14}&$3630\pm70$ & $3.81\pm0.09$ & $4\pm3$\\
TWA 14 & M1.9 & \cite{Herczeg14} & $3500\pm70$ & $4.57\pm0.25$ & $7\pm5$\\
TWA 15 A & M1.5 & \cite{Herczeg14} &$3560\pm140$ & $4.73\pm0.41$ & $10\pm8$\\
TWA 15 B & M2.2 & \cite{Herczeg14} & $3460\pm70$ & $ 4.49\pm0.41$ & $5\pm5$\\
TWA 16 A & M1.8 & \cite{Shkolnik11} & $3520\pm70$ & $ 4.56\pm0.09$ & $8\pm4$\\
TWA 17 & K5 & \cite{Zuckerman04} & $4140\pm160$ & $ 3.21\pm0.18$ & $6\pm3$\\
TWA 18 & M0.5 & \cite{Zuckerman01} & $3700\pm140$ & $3.03\pm0.50$ &$1\pm6$\\
TWA 20 &M3 & \cite{Pecaut13} & $3360\pm100$ &  $5.00\pm0.10$ & $ 10\pm6$ \\
TWA 21 & K3.5 &\cite{Zuckerman04} & $4440\pm210$ & $3.76\pm0.03$ & $19\pm10$\\
TWA 22 A & M6 &\cite{Bonnefoy09} & $2800\pm300$ & $7.39\pm0.03$ & $3\pm3$\\
TWA 22 B & M6 & \cite{Bonnefoy09} & $2800\pm300$ & $7.90\pm0.03$ & $12\pm8$\\
TWA 23 & M3.5& \cite{Herczeg14} & $3260\pm100$ & $5.14\pm0.06$ & $7\pm4$\\
TWA 25 &M0.5& \cite{Herczeg14} & $3700\pm70$ & $3.93\pm0.04$ & $6\pm4$ \\
TWA 26 &M9& \cite{Reid08} & $2380\pm210$ & $9.05\pm0.08$ & $10\pm5$ \\
TWA 27 &M8.25&\cite{Herczeg14} & $2540\pm90$ &$8.78\pm0.05$ & $12\pm6$ \\
TWA 28 & M8.5& \cite{Herczeg14} & $2480\pm90$ &$8.64\pm0.06$ & $9\pm4$ \\
TWA 29 & M9.5& \cite{Looper07} & $2270\pm400$ &$9.31\pm0.36$ & $8\pm7$ \\
TWA 30 A & M5 & \cite{Looper07} & $2880\pm180$ & $5.64\pm0.14$ & $2\pm1$\\
TWA 31 & M4.2 &\cite{Shkolnik11} & $3100\pm130$ & $7.93\pm0.09$ & $49\pm26$\\
TWA 32 &M6.3 &\cite{Shkolnik11} & $2760\pm70$ & $6.02\pm0.09$ & $4\pm2$\\
TWA 33 &M4.7 & \cite{Schneider12} & $2960\pm120$ & $5.86\pm0.19$ & $3\pm2$ \\
2M1025-42 &	M1.0	& \cite{Rodriguez11}	& $3630\pm70$ & $3.65\pm0.29$ & $4\pm2$\\
2M1159-45 & 	M4.5	& \cite{Rodriguez11} & $3020\pm140$ & $4.90\pm0.17$ & $1\pm1$\\
2M1235-41 & 	M3.0 & \cite{Riaz06} & $3360\pm80$ & $6.05\pm0.08$ & $23\pm12$\\
2M1311-42 & 	M1.5 & \cite{Rodriguez11} & $3560\pm 70$ & $4.91\pm0.16$ & $1\pm1$\\
\enddata
\label{tab:ages}
\tablecomments{See Section~\ref{sub:ages} for more details.}
\end{deluxetable}

\begin{table}
\caption{Mean $UVW$ velocities of TWA}
\begin{tabular}{l c c c c}
\hline\hline
Subsample &U & V & W & Mean Age\\
& (km\,s$^{-1}$) & (km\,s$^{-1}$) & (km\,s$^{-1}$) & (Myr)\\
 \hline
All Candidates & $-11.8\pm0.2$ & $-18.9\pm0.2$ & $-5.8\pm0.2$ & $7.9\pm1.0$\\
Distance cut & $-11.7\pm0.2$ & $-18.3\pm0.2$ & $-5.6\pm0.2$ & $7.1\pm0.6$\\
RV cut & $-11.5\pm0.2$ & $-17.5\pm0.1$ & $-5.7\pm0.1$ & $7.1\pm0.8$\\
Parallax cut & $-11.2\pm0.2$ & $-17.7\pm0.1$ & $-5.5\pm0.1$& $7.1\pm0.9$\\
10\,$\sigma$ cut & $-11.5\pm0.2$ & $-17.7\pm0.2$ & $-5.3\pm0.2$& $6.0\pm0.5$\\
\hline
\end{tabular}
\label{tab:meanUVW}
\tablecomments{See Section~6 for more details}
\end{table}

\end{document}